# Investigation of the Solvent-Dependent Photoluminescence Lineshapes in 2,2′-Bithienyl-Substituted 4*H*-1,2,6- Thiadiazin-4-one

Soulianna Kasiouli, Eleni Theodorou, Christalla Kyriakou, Francis Paquin,[#]

Heraklidia A. Ioannidou, Panayiotis A. Koutentis and Sophia C. Hayes*

Department of Chemistry, University of Cyprus, P.O. Box 20537, 1678, Nicosia,

Cyprus

[#]Département de Physique & Regroupement Québécois sur les Matériaux de Pointe,

Université de Montréal, C.P. 6128, Succursale centre-ville, Montréal (Québec),

H3C 3J7, Canada

shayes@ucy.ac.cy

**ABSTRACT**

The solvent dependence of 3,5-bis[(2,2'-bithien)-5-yl]-4*H*-1,2,6-thiadiazin-4-one **BBT** is investigated *via* absorption and photoluminescence spectroscopies. Large differences are observed in the photoluminescence (PL) spectra between polar and non-polar solvents. Narrow bandwidths characterize the PL bandshape in non-polar solvents in contrast to polar solvents, while in CCl$_4$, a significant contribution is observed from a blue-shifted species, possibly formed owing to a reaction between the solvent and the molecule that breaks the $\pi$-conjugation. Variation of the 0-0 band intensity in the PL spectrum as a function of solvent suggests that longer H-aggregates have a larger contribution in the case of polar solvents. Franck-Condon analysis suggests that the solvent dependence of PL bandshape is owed to the presence of two emissive species in polar solvents. We propose that the second emissive species observed in polar solvents could be due to a stabilized charge-transfer state.

## INTRODUCTION

Heteroarenes have been successfully incorporated in π-conjugated oligomers and polymers and a variety including polypyrroles (PPy) and polythiophenes (PT) have found use as the active layer in optoelectronics devices. The latter, especially their milestone alkyl substituted derivatives, are under intensive study as high performance solution-processable electronic polymers. This is due to the tendency of the parent polymer (PT) towards backbone planarity, which can be enhanced by side-chain substitution, and a strong cooperative self-assembly into well-ordered lamellar structures.[1] This behaviour enhances the interchain $p_z$ orbital overlap, thereby improving the electron transport properties (charge mobility), especially in the regioregular polymer samples, making them good candidates for field-effect transistor devices. Heteroarenes with a larger proportion of sulfur and nitrogen atoms, such as thiazole and benzothiadiazole, have been extensively exploited in optoelectronics for their electron-accepting ability. Increased performance of photovoltaic devices was recently found in donor-acceptor systems, where the electron-rich donors were interchanged with electron-poor acceptors in the polymer, as in the case of dioxothienyl-benzothiadiazole copolymers.[2]

4$H$-1,2,6-Thiadiazines are rare heterarenes that like benzo[1,2,5]thiadiazoles host an N-S-N moiety in the skeletal ring. X-Ray studies show the thiadiazine ring to be planar (Fig. 1),[3] and mildly aromatic according to Bird's aromaticity index ($I_A$ 54), (*cf.* $I_A$ = 53 for furan and 100 for benzene).[4] Selected fused 4$H$-1,2,6-thiadiazines have been studied as examples of "extreme quinoids" that have ambiguous aromatic character[5] and several have displayed unusual liquid crystalline properties or behaved as near-infrared dyes.[6] While there are no studies on these 1,2,6-thiadiazines in conjugated polymers, both Woodward[7a] and Rees[7b-d] proposed polymeric thiadiazines as potential organic superconductors. Since 4*H*-1,2,6-thiadiazines are inherently electron poor, their incorporation and evaluation as new acceptor components in narrow-bandgap π-conjugated polymers based on the donor-acceptor approach[2] is rational. Recently developed synthetic methods to attach electron-rich thiophenes to electron-poor 1,2,6-thiadiazines,[8] have enabled the synthesis of π-extended monomers needed for polymerization of donor-acceptor copolymers.[2]

In the present manuscript, we report our investigation of the solvent dependence on the absorption and photoluminescence (PL) spectra of a bis(bithienyl)-thiadiazinone, **BBT**. The PL spectra exhibit interesting solvent dependence, which we interpret within the framework of weakly-coupled H-aggregates using Franck-Condon analysis. Broader bandshapes are observed in solvents more polar than n-hexane suggesting contribution from two emissive species. We propose that the second emissive species is due to a charge-transfer state, which is stabilized through solvent interaction.

**EXPERIMENTAL**

The synthesis of the thienyl-substituted 4*H*-1,2,6-thiadiazine investigated in the present study {3,5-bis[(2,2'-bithien)-5-yl]-4*H*-1,2,6-thiadiazin-4-one, **BBT**} (Fig. 1) has been reported recently.[8a] Dilute solutions of the molecule in n-hexane, acetonitrile ($CH_3CN$), dichloromethane ($CH_2Cl_2$) and carbon tetrachloride ($CCl_4$) (Aldrich) were prepared by dissolving **BBT** (1 mg) in each solvent (20 mL) with subsequent ultrasound treatment (30 min) (sonicator Julabo USR 3). These solutions were further diluted to $10^{-6}$ M. Thin films were made by spin-coating out of a **BBT**/$CH_2Cl_2$/PMMA solution {**BBT** (1 mg) in $CH_2Cl_2$ (20 mL) and 0.1% PMMA [poly(methyl methylacrylate), Aldrich, MW:120 000]}.

The absorption spectra were recorded on a Shimadzu UV-visible 1601 spectrophotometer, with the extinction coefficient ($\varepsilon$) in each solvent determined from a series of solutions of various concentrations (Table I). Photoluminescence (PL) spectra were obtained after excitation with a pulsed 407 nm, 100 ps, 10 MHz diode laser (PicoQuant LDH) and dispersion in a 500 mm spectrograph (SpectraPro2500i, Princeton Instruments) combined with a CCD camera (PIXIS:100F, Princeton Instruments). Excitation spectra were recorded by a JASCO FP-6300 spectrofluorometer. Thin film PL was measured by exciting at 3.06 eV with a 25 mW continuous-wave laser (Newport LQC-405-50-E-00, SpectraPro2300i, Princeton Instrument) modulated at 100 Hz by an optical chopper. PL was sent through a spectrometer (Princeton instrument: SP2300i), detected by a Si photodiode (Electro-optical systems) and analysed by a lock-in amplifier (Standford Research Systems: SR830). The film measurements were done in a cryogenically-cooled cryostat (Cryo Industries of America, CTI 350) kept at 10 K and 290 K. The experiments concerning solutions were performed in air at room temperature and all PL measurements were corrected for the overall spectral response of the instrument.

**RESULTS AND DISCUSSION**

*Photophysical properties.* The absorption spectra of **BBT** (~$10^{-5}$ M) in solvents that range from non-polar to polar are presented in Fig. 2. A blue-shift is observed with an increase in the polarity of the solvent along with slight increase in the vibronic structure (n-hexane and $CCl_4$) of an overall broad and unstructured band. The maximum of the BBT absorption band in $CH_3CN$ is located at 2.65 eV, while the absorption in $CH_2Cl_2$ and $CCl_4$ is shifted to 2.59 eV. Even though the dielectric constant for n-hexane (k = 1.89) as well as the ET(30) value (31 kcal/mol) suggest that this solvent is as nonpolar as $CCl_4$, the spectrum in n-hexane appears more blue-shifted with a maximum at 2.63 eV. The same behavior has been exemplified for **BBT** in cyclohexane (not shown), therefore this effect cannot be attributed to the long solvent chain, but could be ascribed to possible solvation effects (*vide infra*).

Room temperature PL spectra of the above solutions of **BBT** (Fig. 3), show a strong solvent dependence of the PL bandshape. Specifically, the PL spectra of **BBT** in $CH_3CN$ and $CH_2Cl_2$ appear significantly broader than those in the non-polar solvents (n-hexane and $CCl_4$), which demonstrate distinct vibronic structure. The PL maximum for **BBT** in $CH_3CN$ and $CH_2Cl_2$ is located at 2.19 eV, with weak vibronic structure on the red edge of the spectra at 2.09, 1.91 and 1.71 eV. In contrast, a strong vibronic band at 2.38 eV appears in n-hexane, which is assigned to the 0-0 transition. The 0-1 transition is also dominant at 2.23 eV, with two more side bands appearing at 2.07 and 1.9 eV. Distinct vibronic structure is also visible in $CCl_4$ with 0-0 and 0-1 transitions at 2.33 and 2.21 eV, respectively. In addition, in this solvent a broad band at 2.66 eV is very prominent, which we tentatively attribute to shorter conjugated segments produced after reaction of the solvent with the molecule that breaks the conjugation of the chain. In the case of $CH_3CN$ and $CH_2Cl_2$, the transition at 2.19 eV is not assigned to 0-0, but to 0-1 consistent with the assignment in the non-polar solvents; a weak shoulder at 2.35 and 2.33 eV, respectively, is thus identified as the 0-0 transition. This assignment is according to the H-aggregate model, where the 0-0 transition is diminished in ordered aggregation due to symmetry (*vide infra*).

The presence of aggregates is expected to diminish as the concentration of the solution is decreased. Figure 4 presents the concentration dependence of the electronic spectra of **BBT** in polar and non-polar solvents ($CH_3CN$ and n-hexane), where it is clearly observed that in both solvents the absorption and PL spectra remain unchanged with a ten-fold decrease in concentration. Decreasing the concentration further (~$10^{-7}$ M), does not result in essentially any intensity changes in n-hexane, while the slight increase on the blue edge of the PL spectrum in the case of the CH$_3$CN solution could be due to decrease in self-absorption. Therefore, it is possible that the solvent-dependent behavior we observe in the optical spectra was owed to the formation of small aggregates, such as dimers in solution.

Dilute films of $\pi$-conjugated materials in non-conjugated polymer matrices such as PMMA, polyvinyl alcohol (PVA) and poly ethylene oxide (PEO) are usually used to obtain electronic spectra from the isolated molecules.[9, 10] In an attempt to clarify the solvent dependent behavior of **BBT** observed above, films of **BBT** in a 0.1% PMMA matrix were prepared. The solvent used for the solution of **BBT** and PMMA was CH$_2$Cl$_2$. In Figure 5, the PL spectra of **BBT** /PMMA films at 10 K are superimposed on the PL spectrum of **BBT** in CCl$_4$, demonstrating the close resemblance to the PL from **BBT** in a non-polar solvent. It was interesting to note that the red wing of the PL spectrum of **BBT** in CH$_2$Cl$_2$ solution was lost in the PMMA matrix at 10 K, suggesting that two emissive species were present in solutions of polar solvents.

*Theoretical Analysis.* Analysis of the absorption and PL spectra can shed light on the solvent- and concentration-dependent emissive behaviour demonstrated by **BBT**. In non-aggregated single-particle approximation,[11] the bandshape of absorption and emission spectra can reveal the extent of geometry changes upon electronic excitation, which can be extracted from theoretical modelling following a Franck-Condon analysis. In this analysis, the PL vibronic band intensity can be reproduced using Eq. 1:

$$I_{0 \to m} \propto (\hbar\omega)^3 n_f^3 \frac{S^m \exp(-S)}{m!} \qquad (1)$$

where $n_f$ is taken here to be the refractive index of the solution and $m$ denotes the vibrational level. $S$ is the Huang-Rhys factor, which gives a measure of the coupling between the electronic transition and a vibrational mode,[10] thus, this factor is related to the reorganizational energy after excitation. In the case of ordered aggregation, *e.g.* in polymer thin films, where the polymer chains are close to each other and intermolecular interactions are operative, the 0-0 transition from the ground state to the first excited state and *vice versa* is forbidden by symmetry and the oscillator strength for this transition is zero in the absence of disorder. However, in the "weak" excitonic intermolecular coupling limit, the sidebands retain the same relative intensities as in their single-molecule values, but they are slightly diminished consistent with the aggregation-induced reduction in emission quantum yield. In this limit, a disorder-allowed 0-0 component followed by Franck-Condon progression of sidebands can be observed. Therefore, the above expression is modified to reflect a modified Franck-Condon progression with a variable 0-0 amplitude decoupled from the rest of the vibronic progression (represented by α in Eq. 2):[10]

$$I(\omega) \propto (\hbar\omega)^3 n_f^3 e^{-S} \times \left[ \alpha \Gamma(\hbar\omega - E_0) + \sum_{m=1} \frac{S^m}{m!} \Gamma(\hbar\omega - (E_0 - mE_p)) \right] \quad (2)$$

Within this H-aggregate model, the absorption spectra can provide the magnitude of the interchain coupling ($J$), which can be estimated from the ratio of the 0–0 and 0–1 ($S_R$) absorbance peaks. Setting S=1 (Eq. 3),

$$\frac{A_{0-0}}{A_{0-1}} \approx \left( \frac{1 - 0.24 W/E_p}{1 + 0.073 W/E_p} \right)^2 \quad (3)$$

where $E_p$ is the vibrational energy of the main oscillator coupled to the electronic transition and W is the free exciton bandwidth of the aggregates and is equal to 4$J$. A similarly modified Frank-Condon expression as in Eq. 2, which takes into account the effect of aggregation on the relative vibronic intensities, (Eq.4) can be used to fit the absorption spectrum:

$$A \propto \sum_{m=0} \left( \frac{e^{-S} S^m}{m!} \right) \left( 1 - \frac{We^{-S}}{2E_P} G_m \right)^2 \Gamma(\hbar\omega - E_{0-0} - mE_P) \quad (4)$$

where $G_m$ is a constant that depends on $m$ (the vibrational quantum number) according to Eq.5,

$$G_m = \sum_{n(\neq m)} S^n / n!(n-m) \quad (5)$$

Franck-Condon analysis has been employed to interpret the solvent-dependent absorption and PL bandshape of **BBT** and the 0-0 band intensity evolution as a function of solvent in a self-consistent manner (Fig. 6). A standard Franck-Condon progression (Eq. 1) could not adequately reproduce the PL spectra (Fig. 7), even when two modes were included in the calculation. In this latter case, the PL bandshape was calculated using the time-dependent formalism of spectroscopy as applied to fluorescence.[12] Therefore, the modified Franck-Condon approach outlined above, which takes into account the H-aggregate nature in the weak intermolecular coupling regime (Eq. 2), was utilized. The validity of this approximation was verified by the calculation of the free exciton bandwidth for **BBT** (Eq. 3) in n-hexane, $CH_3CN$ and $CH_2Cl_2$, which yielded an appreciably small value (0.013 to 0.018 meV), due to a large 0-0 to 0-1 ratio ($S_R$) that is similar across all solvents. The large value of $S_R$ (~0.93) could suggest that at the concentrations used in this work we are at the isolated molecule limit. However, the inability to fit the PL bandshape in all cases with a normal Franck-Condon expression (Fig. 7) argues that aggregation is significant enough in solution to be taken into account. The absorption spectra of **BBT** in n-hexane, $CH_2Cl_2$ and $CH_3CN$ are shown in Fig. 6(a), (b) and (c) (right) along with the fits obtained with Eq. 4. Similar parameters were used to reproduce the fits as in the PL spectra described below. The calculated value of W is about half the value found from the fit (0.038 meV) (Eq. 4). However, this value, which is small in both cases, is not expected to coincide, since in Eq. 3 the Huang-Rhys parameter is assumed to be 1, while for the fit to the absorption spectrum a value of $S = 1.11$ was used as was obtained from the fit to the PL spectra described below.

Figure 6(a) (left) demonstrates that the PL spectrum of **BBT** in n-hexane, a non polar solvent, can be described entirely by an H-aggregate model, with a vibrational mode at 0.163 eV dominating the coupling to the electronic transition. This vibrational mode (~1290 cm$^{-1}$) does not directly correspond to any of the most intense vibrations in the Resonance Raman spectrum (1435 and 1450 cm$^{-1}$, spectrum not shown), which would be expected to most strongly couple to the electronic transition. However, this is probably due to the missing mode effect (MIME) observed in other conjugated systems,[13] which will be investigated in this system further in a later publication.

As mentioned earlier, in contrast to the PL spectrum in n-hexane, the wide breadth of the PL spectrum of **BBT** in CH$_2$Cl$_2$ and CH$_3$CN cannot be reproduced either by a normal or a modified Franck-Condon expression. Observation of the vibronic structure reveals that the spacing is uneven; specifically the spacing between 0-1 and 0-2 is smaller than the spacing between 0-2 and 0-3 etc, which cannot be explained in terms of anharmonicity. Instead, this suggests the presence of more than one vibronic progression contributing to the observed structure. Assuming the presence of H-aggregates in these solvents as well, we utilized the modified Franck-Condon expression employed in n-hexane to reproduce the blue edge of the spectrum [Fig. 6(b) and (c)]. The emission envelope describing the H-aggregate is characterized by the same vibrational mode as in n-hexane, but with the contribution of the homogeneous and inhomogeneous broadening increasing as would be expected from solvent effects. However, a major difference here is the reduced intensity of the 0-0 component to half the value found for **BBT** in the non-polar solvent, implying increase in the order within the H-aggregate in the more polar solvents. Surprisingly, in contrast to studies involving H-aggregates in thin films of P3HT, a similar decrease in the exciton bandwidth as a function of solvent, as obtained from the absorption spectra, is, however, not observed.

The intensity of the 0-0 band ($I_{0-0}$) has been associated with the inherent disorder ($\sigma$) of H-aggregates, described usually as torsion between subunits in the oligomer or polymer chain.[11] In the elegant work of Spano and coworkers analyzing the PL bandshape of P3HT thin films,[14] an important connection between the theory and experiment was drawn that associated $I_{0-0}$ to the exciton coherence length, *i.e.* the number of molecules over which the center-of-mass of the emitting exciton is coherently delocalized.[15] In the weak excitonic coupling limit and weak intra-aggregate disorder, it was found that $I_{0-0}$ scales as $\left(\frac{1-\beta}{1+\beta}\right)\frac{\sigma^2}{W^2}$, where $\beta = e^{\left(-\frac{1}{l_o}\right)}$ and $l_o$ is the correlation length. Therefore, the coherence length is determined by a competition between the delocalizing influence of excitonic coupling (W) and the localizing nature of disorder ($\sigma$ and a small $l_o$). In addition, this work has demonstrated a dependence of $I_{0-0}$ on the number of chains (*N*) involved in the H-aggregate, with significant change on going from 2 to 10 and 20 chromophores. Theoretical calculations of the structure of the isolated **BBT** (not shown here) demonstrated that the outer thiophene rings on each side of the thiadiazinone are twisted off the plane of the molecule, which can contribute to the disorder in the H-aggregate. These calculations also provided the dipole moment of **BBT** in the ground and excited state, which were small (0.68 and 0.75 D, respectively), most possibly due to the symmetry of the molecule. These dipole moments differ appreciably from the respective values in CH$_3$CN (3.41 D) and CH$_2$Cl$_2$ (1.80 D). Therefore, non-polar solvation could play a key role in the large 0-0 intensity observed in n-hexane compared to the other two solvents that might not be able to adequately solvate the molecule and thus lead to aggregate formation with a larger number of chains involved.

Subtracting the H-aggregate contribution from the PL band, reveals a red-shifted band with distinct vibronic structure that can be equally well reproduced by a normal Frank-Condon (Eq.1) progression and an H-aggregate (Eq.2) (Fig. 8), involving again the same vibrational mode. Interestingly, the difference spectra demonstrate different vibronic intensities, which suggested the extent of structural evolution along the particular vibrational coordinate is varied in the two solvents. In CH$_3$CN, the 0-0 intensity is decreased with respect to the 0-1, while in CH$_2$Cl$_2$, 0-0 and 0-1 appear with similar intensities, which entails that greater structural changes occur in the excited state for the most polar solvent used. However, this could also be a subtraction artifact, and further investigation is necessary. Both spectra demonstrate a broadened red edge that cannot be well reproduced. All the parameters used in the fits are summarized in Tables II, III and IV.

There are two possible explanations for the origin of the red shifted species observed in the more polar solvents. The first could be the generation of a charge-transfer state stabilized by the higher electron affinity solvents (CH$_3$CN and CH$_2$Cl$_2$). The second option is that this red-shifted species is due to a longer aggregate with greater inherent disorder than the blue H-aggregate band, which could be assigned to a dimer. That would explain the fact that we used the same vibronic spacing in the fit, as the vibrational mode belongs to the same molecule but in a different aggregate situation. This could be also an explanation for the fact that the 10 K PL spectrum of **BBT** in PMMA looks like the non-polar solution case (meaning that even at the lowest concentrations there are H-aggregates).

π-Conjugated molecules and polymers with donor-acceptor (D-A) architectures have been known to exhibit intramolecular charge-transfer (ICT) character.[16-18] A widely used criterion to indentify a charge-transfer state is the sensitivity of the UV-vis absorption and PL maximum to the increased polarity of the solvent (solvatochromism).[19-21] Theoretical calculations of the electronic structure of **BBT** in the gas phase (mentioned above and which will be presented in a separate publication) have revealed a charge-transfer transition within the experimentally-observed absorption band of the molecule at 3.12 eV; however the absorption spectrum in Fig. 2 shows little change in the maximum as a function of the polarity of the solvent. This can be justified by the symmetry of the molecule and the proximity of the D and A groups, which leads to small dipole moment values both in the ground and excited state, diminishing thus the effect of solvatochromism. It is therefore possible that a lower-energy charge-transfer excitonic state can be achieved upon photoexcitation. Time-resolved emission spectra (TRES) of **BBT** in $CH_3CN$ (Fig. 9) are key in the clarification of the origin of the red-shifted species. As can be observed in the contour plots, the spectra of **BBT** in n-hexane are narrower than in $CH_3CN$ as expected, with the tail of the latter to expand further to the red. It is apparent that this red tail decays much faster than the blue part of the spectrum, inconsistent with excitonic behaviour. It is known that the emission kinetics of H-aggregated species are slower compared to those of the isolated molecule, precluding thus the explanation that the red-shifted species is due to longer aggregates.[22] Charge-transfer emission kinetics can also be quite slow,[23] however, we believe that the proximity of the D and A groups must contribute to the faster kinetics observed.

More support for the assignment of the red-shifted species to a charge-transfer state was provided from evidence for the generation of charged photoexcitations in oligomeric thiophenes. [24] Interestingly, the photogeneration of the $3T^+$ cation radical the solvent participated actively in the process. Specifically, the signal associated with this cationic radical was only observed in chlorinated solvents and $CH_3CN$/water, but not in benzene, cyclohexane, or diethyl ether.[25] Crucial to the formation of this charged state was the presence of a suitable electron acceptor. Therefore, in the case of the thienyl-substituted thiadiazine investigated in the present study, the presence of the central thiadiazinone as the electron acceptor along with the use of $CH_3CN$ and $CH_2Cl_2$ collaborate to the formation of a charged species in the low energy side of the PL band (~ 1.7-1.8 eV) most possibly located on the thiophene chains. Such low energy band was observed for $3T^+$ at 1.46 eV. Direct observation of this cationic radical species can be obtained through vibrational spectroscopy, and experiments are underway.

**CONCLUSIONS**

In the present study we have shown that the emission and absorption spectra of **BBT** in n-hexane, $CH_2Cl_2$ and $CH_3CN$ are consistent with the presence of weakly-coupled H-aggregates. In solvents more polar than n-hexane, a second red-shifted emissive species is apparent, which has been tentatively assigned to a low energy charge transfer-state stabilized by the solvent. The vibronic structure of the PL band attributed to this species varies from $CH_2Cl_2$ to $CH_3CN$, with greater conformational relaxation observed in the most polar solvent used.

**ACKNOWLEDGMENTS**

The authors are thankful to the Optoelectronics Group at the University of Cambridge for the use of their PL instrument and particularly to Dr. Jenny Clark for helpful discussions. Helpful discussions with Prof. Carlos Silva, Prof. Frank Spano and Prof. Franco Cacciali are greatly acknowledged. The authors would also like to thank Dr. Christophe Jouvet for preliminary theoretical calculations on **BBT**. Finally, the authors thank the Cyprus Research Promotion Foundation for funding (grant no. ΠΕΝΕΚ/ΕΝΙΣΧ/0504/08).

**Table I**.  Optical properties of **BBT** in various solvents.

| Solvents | $\lambda_{max, abs}$ (nm) | $\lambda_{max, em}$ (nm) | $\varepsilon$ (cm$^{-1}$M$^{-1}$) |
|:---:|:---:|:---:|:---:|
| CH$_3$CN | 467 | 566 | 33731 |
| n-hexane | 471 | 556 | 23557 |
| CCl$_4$ | 478 | 561 | 28567 |
| CH$_2$Cl$_2$ | 478 | 566 | 27477 |

**Table II:** Parameters used in the simulation of the absorption and photoluminescence spectra based on an H-aggregate model.

| Absorption | | | | | | |
|---|---|---|---|---|---|---|
| Solvent | $de^a$ (eV) | $S^b$ | $W^c$ | $\Gamma_{hom}^d$ (eV) | $\Gamma_{inhom}^e$ (eV) | $E_{00}^f$ (eV) |
| n-hexane | 0.160 | 1.11 | 0.0379 | 0.150 | 0.095 | 2.49 |
| $CH_2Cl_2$ | 0.160 | 1.11 | 0.0385 | 0.160 | 0.110 | 2.45 |
| $CH_3CN$ | 0.160 | 1.11 | 0.0385 | 0.160 | 0.110 | 2.50 |
| Photoluminescence | | | | | | |
| Solvent | $de$ (eV) | $S$ | $\alpha$ | $\Gamma_{hom}$ (eV) | $\Gamma_{inhom}$ (eV) | $E_{00}$ (eV) |
| n-hexane | 0.163 | 1.15 | 0.8 | 0.110 | 0.096 | 2.385 |
| $CH_2Cl_2$ | 0.160 | 1 | 0.42 | 0.125 | 0.12 | 2.35 |
| $CH_3CN$ | 0.163 | 1 | 0.48 | 0.120 | 0.12 | 2.32 |

[a]: the vibrational energy of the C=C symmetric stretch ($\hbar\omega$)

[b]: the Huang-Rhys factor

[c]: exciton bandwidth

[d]: homogeneous broadening (in eV)

[e]: inhomogeneous broadening (in eV)

[f]: 0-0 transition energy

[g]: a constant whose amplitude varies during the fit. Is a strong function of the disorder width and spatial correlation length.

**Table III:** Parameters used in the simulation of the absorption and photoluminescence spectra based on a normal Franck-Condon progression.

| Photoluminescence | | | | | | |
|---|---|---|---|---|---|---|
| **Solvent** | **de (eV)** | **S** | **α** | **$\Gamma_{hom}$ (eV)** | **$\Gamma_{inhom}$ (eV)** | **$E_{00}^{f}$ (eV)** |
| n-hexane | 0.163 | 1.45 | 0 | 0.11 | 0.096 | 2.38 |
| $CH_2Cl_2$ | 0.163 | 1.7 | 0 | 0.125 | 0.12 | 2.29 |
| $CH_3CN$ | 0.163 | 2 | 0 | 0.125 | 0.12 | 2.31 |

**Table IV:** Parameters used in the simulation of the difference photoluminescence spectra based on standard Franck-Condon progression and H-aggregate model.

| Difference Photoluminescence spectra | | | | | | |
|---|---|---|---|---|---|---|
| **Solvent** | **de (eV)** | **S** | **α** | **$\Gamma_{hom}$ (eV)** | **$\Gamma_{inhom}$ (eV)** | **$E_{00}$ (eV)** |
| $CH_2Cl_2$ H-aggregate model | 0.16 | 1.06 | 0.91 | 0.105 | 0.103 | 2.08 |
| $CH_2Cl_2$ Normal FC model | 0.16 | 1.34 | 0 | 0.105 | 0.103 | 2.08 |
| $CH_3CN$ H-aggregate model | 0.163 | 1.6 | 0.73 | 0.11 | 0.1 | 2.067 |
| $CH_3CN$ Normal FC model | 0.163 | 1.67 | 0 | 0.11 | 0.1 | 2.067 |

**FIGURE LEGENDS**

Figure 1. Chemical structure of 3,5-bis[(2,2'-bithien)-5-yl]-4*H*-1,2,6-thiadiazin-4-one, (**BBT**).

Figure 2. Normalized absorption spectra of **BBT** in various solvents. The concentration of **BBT** in solution was on the order of $10^{-5}$ M. ($CH_3CN$: $6.23\times10^{-5}$ M, n-hexane: $8.4\times10^{-5}$ M, $CH_2Cl_2$: $1.9\times10^{-4}$ M, $CCl_4$: $9.7\times10^{-5}$ M).

Figure 3. Normalized photoluminescence spectra of **BBT** solutions ($10^{-5}$ M). ($CH_3CN$: $6.23\times10^{-5}$ M, n-hexane: $8.4\times10^{-5}$ M, $CH_2Cl_2$: $1.9\times10^{-4}$ M, $CCl_4$: $9.7\times10^{-5}$ M).

Figure 4. Normalized absorption and photoluminescence spectra of **BBT** solutions in (a) $CH_3CN$ and (b) n-hexane at different concentrations. (black) $10^{-5}$ M (red) $10^{-6}$ M, and $10^{-7}$ M (blue circles). Spectra normalized with respect to the 0-1 transition.

Figure 5. Normalized PL spectra of **BBT** in $CCl_4$ solution (black line) and $CH_2Cl_2$//PMMA film (red line) at 10 K.

Figure 6. Left: PL spectra (blue line) of **BBT** in n-hexane (a), $CH_2Cl_2$ (b), and $CH_3CN$ (c). The red line displays the PL spectra of the H-aggregate obtained via Eq. 2. The black line is the difference between the measured PL spectra and the modelled aggregate spectrum. (b) and (c). Right: Absorption spectra (red line) of **BBT** in n-hexane (a), $CH_2Cl_2$ (b) and $CH_3CN$ (c). The blue line displays the PL spectra of the H-aggregate using Eq. 4.

Figure 7. PL spectra of **BBT** in (a) n-hexane (red line), (b) $CH_2Cl_2$, and (c) $CH_3CN$. The blue line displays the calculated PL spectra reproduced with Eq. 1.

Figure 8. (a) Normalized difference PL spectra of **BBT** in $CH_3CN$ (red line) and $CH_2Cl_2$ (blue line). Franck-Condon fit of the difference spectra for a normal progression (black line) and an H-aggregate (red line) for **BBT** in $CH_3CN$ (b) and $CH_2Cl_2$ (c).

Figure 9. Contour plot of time-resolved emission spectra of **BBT** in $CH_3CN$ (top) and n-hexane (bottom).

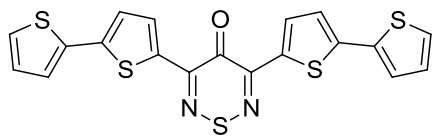

**BBT**

Figure 1

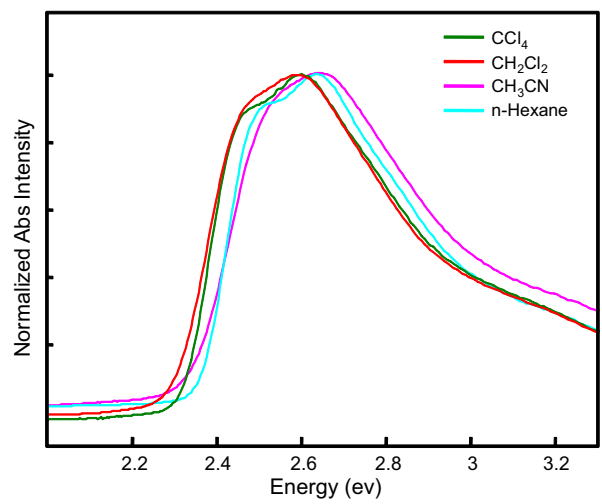

Figure 2.

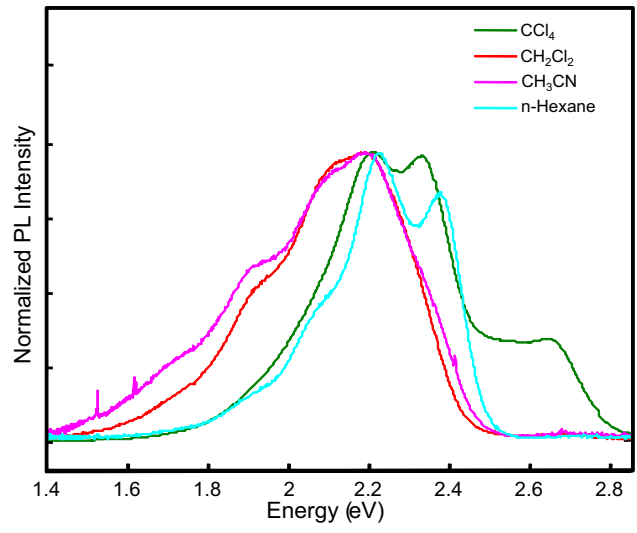

Figure 3.

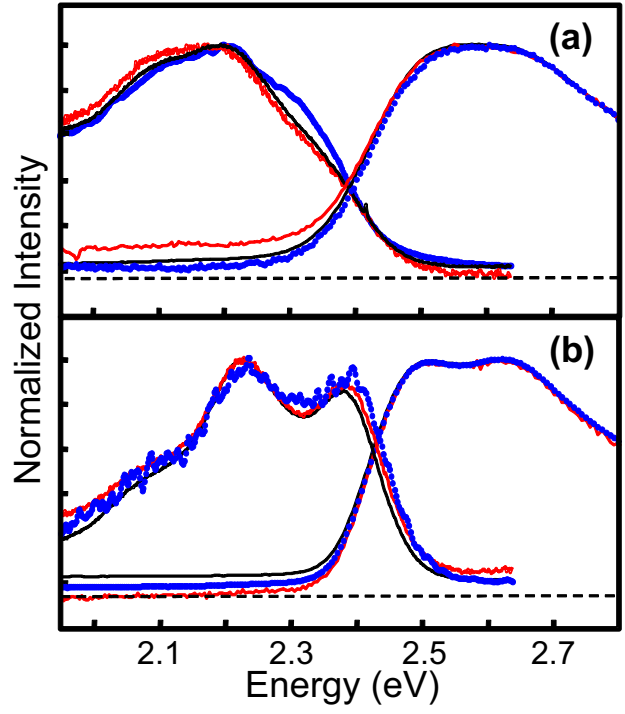

Figure 4.

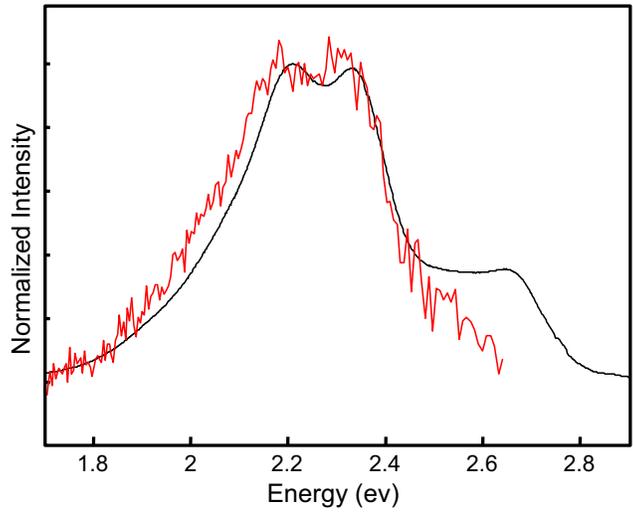

Figure 5.

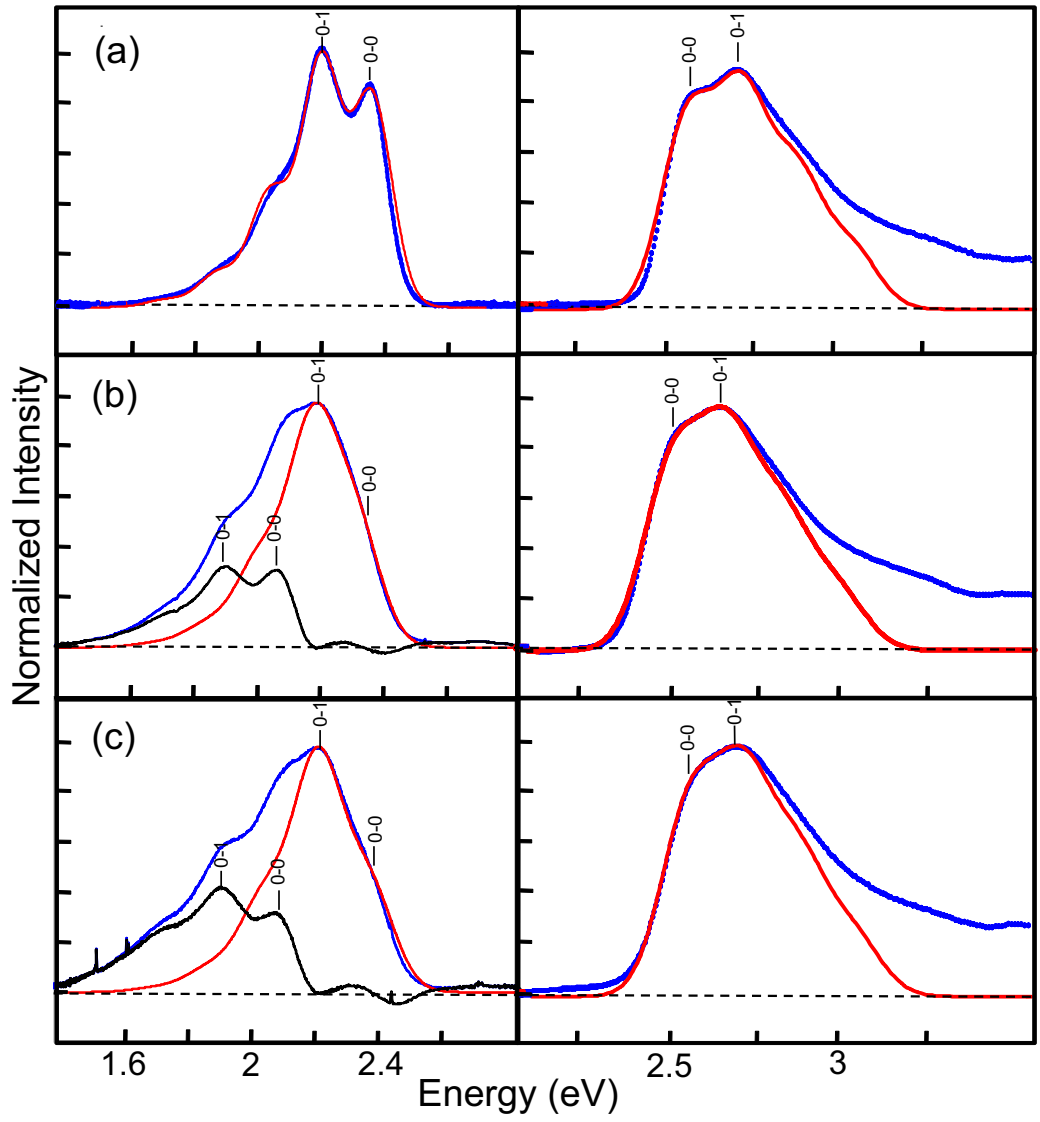

Figure 6.

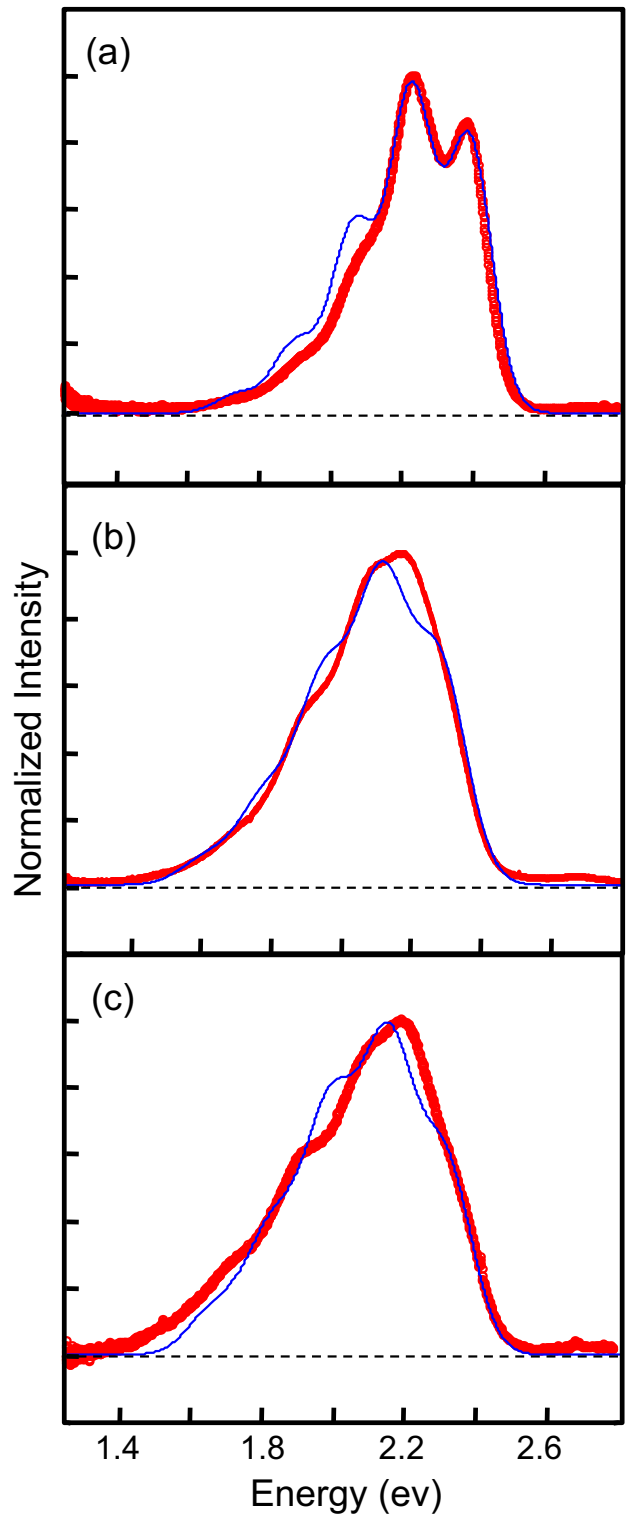

Figure 7.

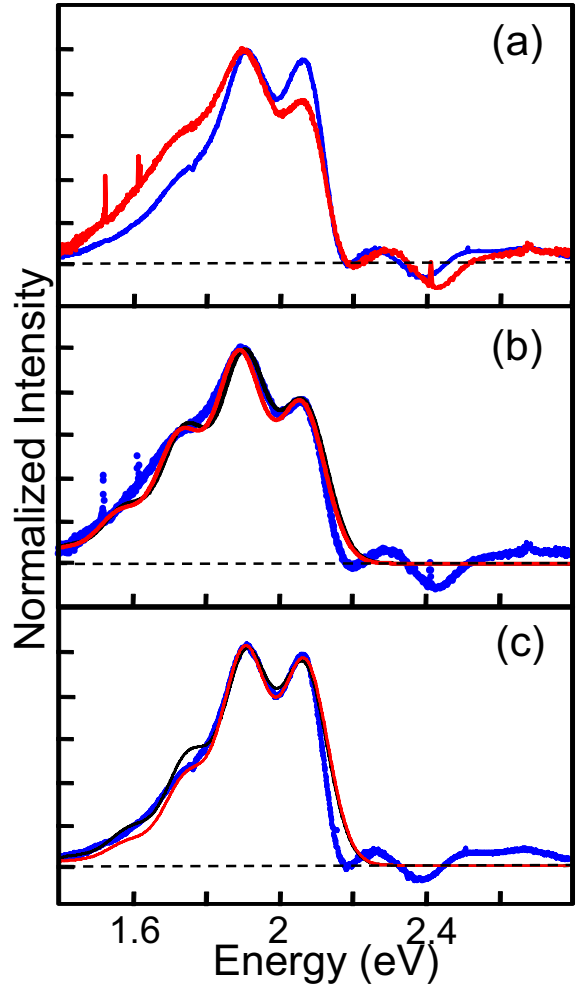

Figure 8.

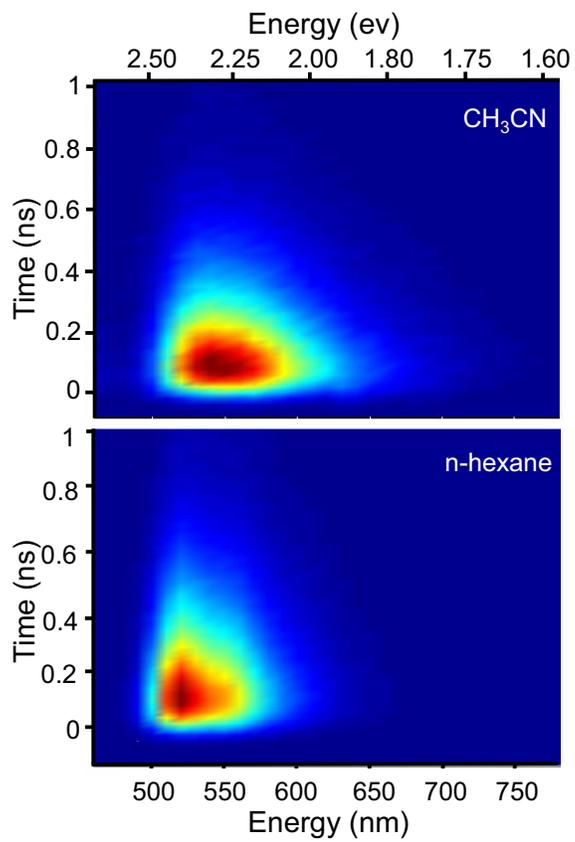

Figure 9.